\documentclass[twocolumn]{aastex62} 
\newcommand{\eg}{e.g.,}

\usepackage{amsmath}

\usepackage{appendix}
\usepackage{float}
\usepackage{savesym}
\savesymbol{tablenum}
\usepackage{siunitx}
\restoresymbol{SIX}{tablenum}
\submitjournal{ApJ}
\DeclareSIUnit\mearth{M_\oplus}

\begin{document}

\title{Realistic On-The-Fly Outcomes of Planetary Collisions: Machine Learning Applied to Simulations of Giant Impacts}

\correspondingauthor{Saverio Cambioni}
\email{cambioni@lpl.arizona.edu}

\author[0000-0001-6294-4523]{Saverio Cambioni}
\affil{Lunar and Planetary Laboratory, University of Arizona, 1629 E. University Blvd., Tucson, AZ 85721, USA}

\author[0000-0003-1002-2038]{Erik Asphaug}
\affil{Lunar and Planetary Laboratory, University of Arizona, 1629 E. University Blvd., Tucson, AZ 85721, USA}

\author[0000-0002-8811-1914]{Alexandre Emsenhuber}
\affil{Lunar and Planetary Laboratory, University of Arizona, 1629 E. University Blvd., Tucson, AZ 85721, USA}

\author[0000-0002-9767-4153]{Travis S.\ J.\ Gabriel}
\affil{School of Earth and Space Exploration, Arizona State University, 781 E. Terrace Mall, Tempe, AZ 85287, USA}

\author{Roberto Furfaro}
\affil{Systems and Industrial Engineering Department, University of Arizona, 1127 E. James E. Rogers Way, Tucson, AZ 85721, USA}

\author[0000-0001-5475-9379]{Stephen R.\ Schwartz}
\affil{Lunar and Planetary Laboratory, University of Arizona, 1629 E. University Blvd., Tucson, AZ 85721, USA}

\begin{abstract}
Planet formation simulations are capable of directly integrating the evolution of hundreds to thousands of planetary embryos and planetesimals, as they accrete pairwise to become planets. In principle such investigations allow us to better understand the final configuration and geochemistry of the terrestrial planets, as well as to place our solar system in the context of other exosolar systems. These simulations, however, classically prescribe collisions to result in perfect mergers, but computational advances have begun to allow for more complex outcomes to be implemented. Here we apply machine learning to a large but sparse database of giant impact studies, streamlining simulations into a classifier of collision outcomes and a regressor of accretion efficiency. The classifier maps a 4-Dimensional parameter space (target mass, projectile-to-target mass ratio, impact velocity, impact angle) into the four major collision types: merger, ``graze-and-merge'', ``hit-and-run'', and disruption. The definition of the four regimes and their boundary is fully data-driven; the results do not suffer from any model assumption in the fitting. The classifier maps the structure of the parameter space and provides insights about the outcome regimes. The regressor is a neural network which is trained to closely mimic the functional relationship between the 4-D space of collision parameters, and a real-variable outcome, the mass of the largest remnant. This work is a prototype of a more complete surrogate model, based on extended sets of simulations (``big data''), that will quickly and reliably predict specific collision outcomes for use in realistic $N$-body dynamical studies of planetary formation.  
\end{abstract}


\section{Introduction}
\label{intro}

The idea of giant impacts has gone well beyond Moon formation \citep[\eg][]{1975Icar...24..504H,1987AREPS..15..271S,1989Icar...81..113B,2001Natur.412..708C} to give new understandings of planet formation during the ``late stage'', when bodies that are similar in size collide at one to several times their mutual escape velocity $v_{\rm esc}$,
\begin{equation}
\label{esc}
    v_\mathrm{esc}=\sqrt{\frac{2G(M_\mathrm{T}+M_\mathrm{P})}{R_\mathrm{coll}}},
\end{equation}
\noindent
where $M_\mathrm{T}$ is the mass of the target, $M_\mathrm{P}$ is the mass of the projectile and $R_\mathrm{coll}=R_\mathrm{T}+R_\mathrm{P}$ is the separation at initial contact \citep[e.g.,][]{1985Sci...228..877W,2010ChEG...70..199A}. Supported by increasingly-sophisticated models, hypotheses have emerged for the giant impact formation of planets including Mercury \citep{2007SSRv..132..189B,2014NatGe...7..564A,2018ChauMercury}, Pluto-Charon \citep[\eg][]{2005Sci...307..546C,2011AJ....141...35C}, Haumea \citep{2010ApJ...714.1789L}, Titan \citep{2013Icar..223..544A}, and the Moon. In a broad sense, giant impact events have had a significant role in determining the final physical properties of rocky/icy planets. 

In \textit{N}-body dynamical studies, planetary embryos orbit the Sun and each giant impact is typically assumed to be fully accretionary, so that \textit{N} only decreases in time. However, perfect merging is known \citep{chambers2013} to be a problematic oversimplification of more complex outcomes, as has been demonstrated by decades of detailed hydrocode simulations \citep[\eg][]{2006Natur.439..155A} using methods such as Smoothed-Particle Hydrodynamics (SPH) described below. The most common collision events at the end-stage of terrestrial planet formation in the solar system involve similar-sized bodies and $v_\mathrm{coll}/v_\mathrm{esc}=1 - 4$ \citep{1999Agnor}. Over this range of mass ratios and impact velocities, collision outcomes span all the regimes of accretion, erosion, and ``hit-and-run'' \citep{2012ApJ...745...79L}. Some more advanced \textit{N}-body approaches have implemented simple rules for limiting accretion efficiency \citep[\eg][]{chambers2013}, but approximations such as perfect mergers are still the norm \citep[\eg][]{Obrien2006,raymond2009}. 

One ultimate strategy is to model collisions ``on the fly'', \eg\ using SPH to model a given impact event while the $N$-body evolution is in progress \citep[\eg][]{2017DPS....4950802H}. But in practice this has limitations. In order for a giant impact simulation to run in less than an hour, a practical limit when the $N$-body evolution must wait, the hydrocode resolution is limited to a $\sim\num{e4}$ particles, which is only adequate for classifying the most basic outcomes \citep{2004ApJ...613L.157A}. Then there is the concern of data reduction. Well-resolved giant impact simulations often generate multiple debris products (including intact remnants), and these must be identified and characterized in each output file to be fed back into the $N$-body code. These include the projectile ``runner'' in case of ``hit-and-run'' \citep{2006Natur.439..155A}, and other self-gravitating clumps and debris. ``Graze-and-merge'' collisions can spin off escaping bodies up to a third the size of the progenitors \citep{2013Icar..223..544A}, and head-on impacts appear to make fields of sizable clumps \citep[\eg][]{2018Sugiura}. Keeping track of all this requires post-processing analysis of the collision outcome and increases $N$, which can stall the evolution. Ensuring convergence of the debris field requires larger numbers of particles than a nominal simulation \citep[\eg][]{2015genda}. 

However, a detailed description of the debris field is neither needed nor desired. Instead, we would prefer a summary description of the two or three major bodies emerging from the giant impact, their thermodynamic and orbital dynamic states, and useful statistics regarding the remaining debris, e.g., their characteristic sizes and velocity distributions, as well as the overall mass, momentum and composition. Lastly, it must be recognized that the knowledge of the specific impact properties (angle of impact relative to spin state of planet, for example) are in fact completely unknown, so that running a superb 3-D simulation of a specified giant impact is misplaced effort, unless results can be generalized in some way. Our approach is to use high-fidelity SPH calculations as a training dataset, beginning with the impact simulations published by \cite{2011PhDReufer} that is also the basis for \cite{2019ApJGabriel}, who develop a forward-functional model from the same dataset.

We use SPH to model giant impacts on planetary bodies such as the Moon, Mercury and Mars \citep[e.g.,][]{2014NatGe...7..564A, 2015aste.book..661A, 2012Icar..221..296R}. Each SPH outcome is a complex $N$-Dimensional state (consolidated planets, clumps, unconsolidated ejecta, and their thermodynamic states and other characteristics) that requires detailed analysis. Giant impacts cover a large range of input parameters, and are intrinsically three-dimensional events. As an example, a colliding pair of planets is represented by masses $M_1$ and $M_2$, their impact velocity and angle, target and impactor spin rate and orientation, plus some assumptions on their composition and internal structure. Performing 5 realizations of each variable would require nearly \num{400000} simulations, just to produce a coarse mapping of the parameter space. The necessity of a detailed coverage of the parameter space couples with the requirement of precise (high-resolution) simulations. Simulations with $10^6$ particles have become standard \citep[e.g.,][]{2013Icar..222..200C,2017ApJ...845..125H,2018Icar..301..247E}, and runs are extended to many gravitational times $\tau_g=\sqrt{4\pi/3\rho G}$ \citep{2017A&A...597A..62J}.

We apply machine learning to build an accurate data-driven model of giant impacts, which does not simply interpolate the available data, but rather generalizes the underlying functional relationship between impact properties and collision outcomes. It does so as to fit the available data, but not to over-fit it; that is, to be inclusive of the expectation of new data that is yet to be observed. The data described below is ideal for an initial study, but is being superceded by much higher fidelity models; one of the advantages of this approach is that higher fidelity data can be added to lower fidelity data in a weighted manner as they become available. 

We present two distinct machine-learned response functions for collisions in the gravity regime: a \textit{classifier} of collision types and a \textit{regressor} of accretion efficiency. These functional models -- compact algorithms -- map the outcome of a giant impact (post-collision end state) into a 4-Dimensional parameter space, i.e., mass of target, projectile-to-target mass ratio, impact velocity and impact angle.  The training is performed on existing giant impact simulations between similar-size differentiated chondritic bodies. The resulting surrogate collision models give a reliable result to within a known degree of confidence and at a highly reduced computational time (with respect to full giant impact simulations, i.e., on the order of seconds on a single computing thread). Therefore, they are designed to apply especially well to \textit{N}-body evolution calculations and to constrain pre-impact dynamical conditions from hypothesized post-collision scenario \citep{2018MNRAS.474.2924J}.

The paper is organized as follows. In Section~\ref{dataset} we describe the available dataset. In Section~\ref{machine_learning} we provide an introduction to machine learning, with a focus on the two distinct algorithms used to train the classifier of collision types and the regressor of accretion efficiency: Support Vector Machine (Section~\ref{SVM}) and Neural Network (Section~\ref{NN}), respectively. In Sections \ref{sec:result} and \ref{discussion} we present and discuss the predictions by the trained algorithms regarding the post-collision end states and the characterization of the parameter space. Finally, we discuss the potentialities of the methodology and future work/application in Section~\ref{concl}.

\section{Materials and Methods}

\subsection{Dataset}
\label{dataset}
\begin{figure}
    \centering
    \includegraphics[width=\linewidth]{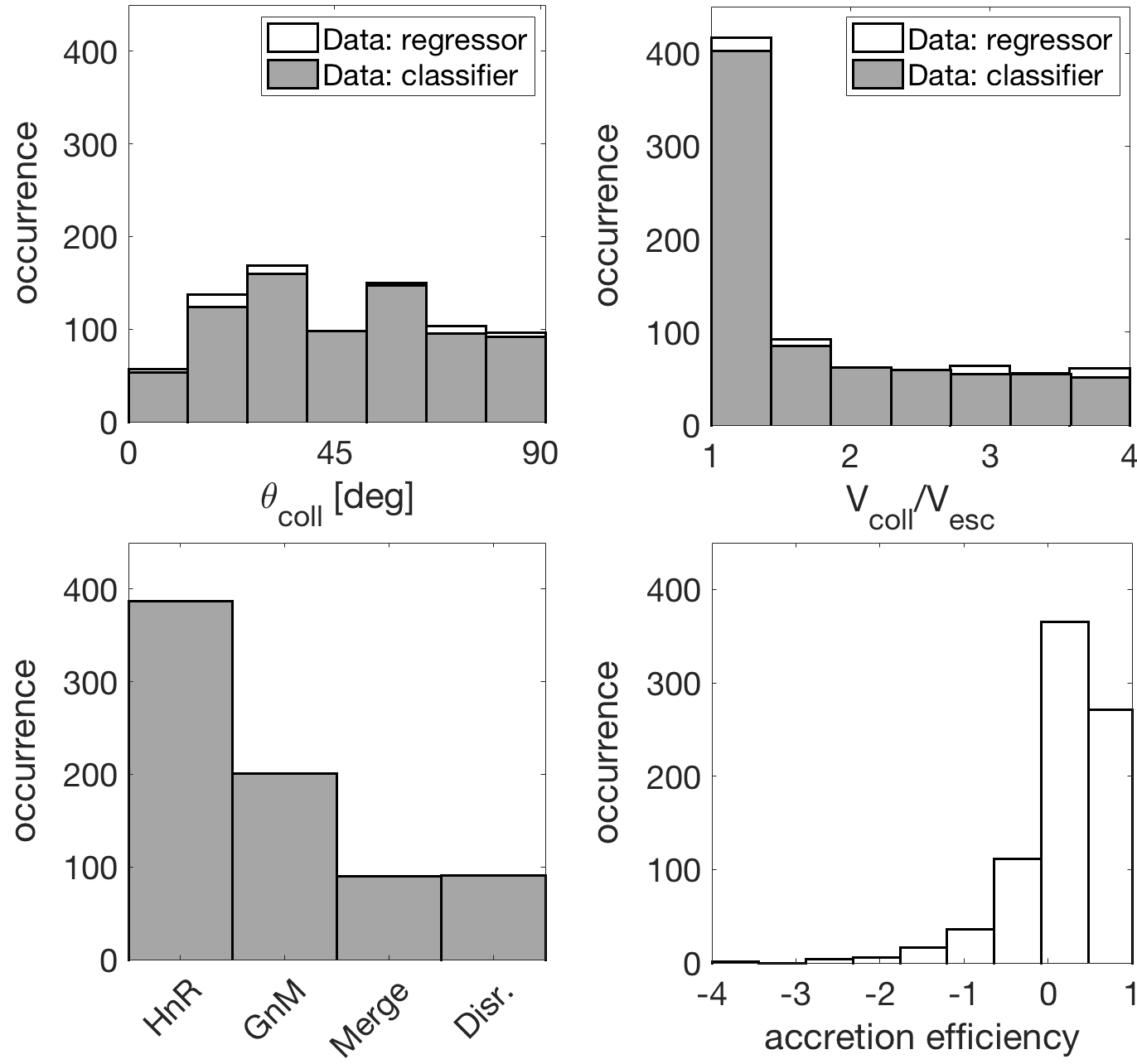}
    \caption{Top-left and top-right panels: frequency distributions of input impact angle $\theta_{coll}$ and velocity $v_{coll}/v_{esc}$, respectively; the values for the other two input parameters that are present in the dataset -- target mass and mass ratio (projectile/target) -- are provided in Table~\ref{tab:dataset}. Bottom-left panel: frequency distribution of the collision classes as labelled in the classification task (Sections \ref{SVM} and \ref{class}); on the x-axis, ``HnR'' refers to the simulations labelled as ``hit-and-run'' cases, ``GnM'' refers to the ``graze-and-merge'' cases, and ``Disr.'' refers to the ``disruption'' cases. Bottom-right panel: frequency distribution of accretion efficiency values -- Equation \ref{acceff} -- which are used in the regression task (Sections \ref{NN} and \ref{surr_mod}). Simulations from \citet{2011PhDReufer}. Details on the database of simulations and detailed physical analysis are provided in \citet{2019ApJGabriel}.}
    \label{histo_data}
\end{figure}

In this study, we use SPH simulations from \citet{2011PhDReufer}. Completed at $\sim\num{200000}$-particle resolution, the dataset spans a wide range of parameters: target mass, mass ratio (projectile/target), impact angle, and impact velocity. The bodies are similar in size and initially non-rotating. They are differentiated with a chondritic composition of 30\% iron and 70\% silicate. The values for the first two parameters that are present in the dataset are provided in Table~\ref{tab:dataset}. For each pair, more than 100 runs with different impact velocity and angle are performed, ranging between 0 and 90$^\circ$, and 1 to 4 times the mutual escape velocity, respectively (top-left and top-right panels of Figure \ref{histo_data}). These conditions are the most relevant to the late-stage planet formation \citep{2012ApJ...751...32S,chambers2013}.  Among the impact parameters, we do not include the initial spin rates, which require three additional variables for each of the bodies (one for the magnitude and two for the orientation). 
The target and impactor spin rates, however, have been found to be relevant for the overall impact outcome \citep[e.g.,][]{2005Sci...307..546C,2011AJ....141...35C}; we intend to include this parameter in future machine learning applications.

The collisions in our dataset are modeled using the SPH technique. SPH is a physically-based hydro-dynamical model that uses a Lagragian description, which is suited for collision modeling, where a large range of densities is expected; no grid is required, in contrast with Eulerian methods. Quantities are obtained by interpolating over $\sim 50-100$ neighbor particles using a kernel function -- in this case a $\beta$-spline \citep{1985A&A...149..135M}. Spatial derivatives are retrieved using the derivative of the kernel, so that no grid is required. Time evolution is provided by Euler's equations: mass conservation to obtain the density, energy conservation for the internal energy, and momentum conservation with pressure gradient and self-gravity. An artificial viscosity is added to resolve shocks as is common in nearly all SPH implementations in planetary science \citep[\eg][]{1992ARA&A..30..543M}. An equation of state is required to obtain the pressure from the density and internal energy; we use M-ANEOS for SiO\textsubscript{2} and ANEOS for iron \citep{2007M&PS...42.2079M, ANEOS} -- a common choice for such studies. Self-gravity is based on a hierarchical spatial tree \citep{1986Natur.324..446B}, where contributions from distant regions are estimated using a multi-pole approximation. The same tree is used to walk the nearest-neighbor search, a process that occurs throughout the simulation.

\begin{figure}
    \centering
    \includegraphics[width=\linewidth]{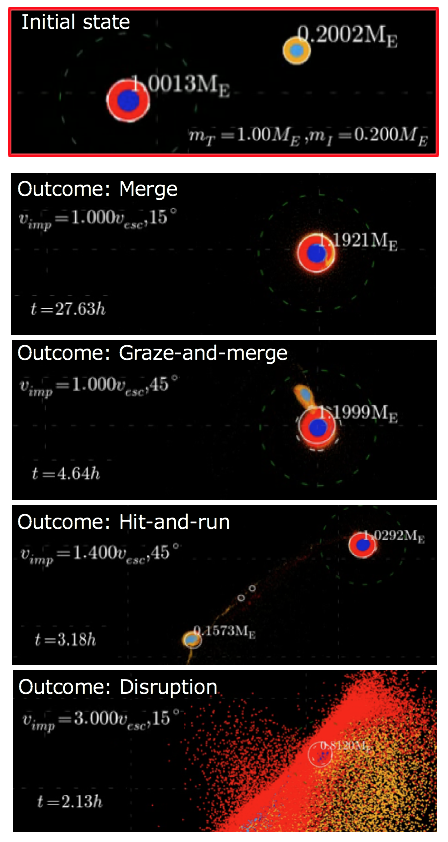}
        \caption{Different combinations of the 4 impact properties (predictors: mass of the target, projectile-to-target mass ratio, impact angle, impact velocity) lead to different collision outcomes (responses). The SPH codes allow easy visualization of the results in form of short clips. As an example, the top panel of the figure shows the initial state of the simulation (target and impactor before the collision). The other panels show the collision type for various impact velocities and angles. From top to the bottom: a merging event, resulting from a head-on collision at low impact velocity; a ``graze-and-merge'' event, resulting from a collision at the most probable impact angle \citep[\SI{45}{\degree},][]{1962BookShoemaker} and low impact velocity; a ``hit-and-run'' event, resulting from a collision at an impact angle of \SI{45}{\degree} and moderate impact velocity; a disruptive event, resulting from a head-on collision at high impact velocity. The time after the collision (in hours) is reported at the bottom left of each frame. Simulations from \citet{2011PhDReufer}. Details on the database of simulations and detailed physical analysis are provided in \citet{2019ApJGabriel}.}
        \label{movie_nigth}
\end{figure}

Each simulation begins with the bodies approaching from several radii away, to allow for tidal deformation to take place prior to the collision. The initial conditions are determined assuming a two-body problem so that the velocity and angle at initial contact follow the prescribed values. The simulations are evolved for $50\tau_\mathrm{coll}$ past initial contact, with $\tau_\mathrm{coll}$ being the collision time scale defined as
\begin{equation}
\label{tg}
    \tau_\mathrm{coll} = \frac{2R_\mathrm{coll}}{v_\mathrm{coll}},
\end{equation}
\noindent
where $v_\mathrm{coll}$ is the impact velocity and $R_\mathrm{coll}=R_\mathrm{T}+R_\mathrm{P}$ is the separation at initial contact. The indexes T and P refer to the target and projectile respectively. Once the simulation has finished, the resulting bodies are found using the following iterative algorithm: particles pairs are iterated over, starting with the ones that have the lowest gravitational potential energy, and checked whether the pair is bound. If a pair is bound, then a new clump is started, and the iteration continues checking particles against the new clump. For each particle added, the iteration is repeated until no further particle is found to be bound to the clump. This procedure is also used to compute the mass of the largest remnant of the collisions. Details on the database of simulations and detailed physical analysis are provided in \citet{2019ApJGabriel}. Snapshots of the movie rendering of these simulations are shown in Figure~\ref{movie_nigth}.

\begin{table}
    \centering
    \begin{tabular}{|c|c|}
    \hline
        Target mass [$\mathrm{M_\oplus}$] & Mass ratio (projectile/target)\\
        \hline
        $1$ & 0.20, 0.70 \\
        $10^{-1}$ & 0.10, 0.20, 0.35, 0.70 \\
        $10^{-2}$ & 0.20, 0.70 \\
        \hline
    \end{tabular}
    \caption{Pairs of target masses and projectile-to-target mass ratio present in the collisions dataset from \citet{2011PhDReufer,2019ApJGabriel}. For each pair, more than 100 runs with different impact velocity and angle are performed, ranging between 0 and 90$^\circ$, and 1 to 4 times the mutual escape velocity, respectively (Figure \ref{histo_data}). }
    \label{tab:dataset}
\end{table}

The simulations in the dataset use SPH in the original, fluid mode, where the equation of motion is derived only from the pressure gradient \citep[\eg][]{1992ARA&A..30..543M} and self-gravity. This is appropriate when the stresses of gravity exceed the possible mechanical strengths, and for this reason the existing dataset has its lower limit at \SI{1400}{\kilo\meter} diameters (the so-called ``gravity regime''). For bodies 100--\SI{1000}{\kilo\meter} in diameter, it has been shown that friction \citep[e.g.][]{2015P&SS..107....3J} and strength \citep{2018Icar..301..247E} are important, with the potential to challenge our ideas for the origin of moons and embryos during the late stage of planet formation. For super-Earth and Neptune-mass bodies (\SI{10000}{\kilo\meter} and larger), the dominant variables are thermodynamic processes, shocks and gravity \citep{2009Marcus,2010MarcusB,2015ApJ...812..164L,2018LPI....49.1886K}. For this work we limit ourselves to Earth-sized planets and smaller, because a sufficiently large database for super-Earth and Neptune-sized collisions is not reported in the literature.

\subsection{Machine Learning}
\label{machine_learning}
Machine Learning (ML) is a subfield of data analysis that lies at the cornerstone between statistical methods and computer science as well as at the core of artificial intelligence. Originally conceived to address the question of how to build computers that can autonomously improve through direct experience, ML enables machines to learn features and trends from the available data. Over the past few years, encouraged by advancements in parallel computing technologies (e.g., Graphic Processing Units, GPUs), availability of massive labeled data as well as breakthrough in understanding of deep neural networks, there has been an explosion of ML algorithms that can accurately process images for classification and regression tasks, \eg\ image and video recognition \citep{krizhevsky2012imagenet}, natural language processing \citep{socher2012deep}, speech recognition \citep{hinton2012deep}. State-of-the-art ML techniques allow for several advantages: they can streamline the generation of data sets to most efficiently explore regions of interest in a large parameter space; and they can perform accurate mappings of initial conditions and end-states, with associated probabilities, taking into account a high-dimensional parameter space. This is in contrast to human operators that are often limited to a mostly 2-D understanding of the data. ML schemes take advantage of this `big data problem' to spot new and sometimes unexpected correlations.

ML techniques are divided into supervised (or predictive) and unsupervised (or descriptive) methods. Supervised methods rely on a training set of data, both with features/predictors and labels, that is known with some level of confidence. As an example, in a giant impact, a set of predictors (e.g., impact angle, impact velocity, mass of the target) results into a collision outcome, such as merger or disruption (the label). In supervised learning, the dataset is split into training samples, validation samples (data used to measure generalization capability of the algorithm), and testing samples (data that do not affect training and are used as an independent measure of performance during and after training). By contrast, unsupervised methods do not label the data directly into classes but rather attempt to find patterns and trends underlying in the data. Usually, such algorithms (\eg\ K-mean, \citealp{ahmad2007k}) require an initial assumption on the data (\eg\ number of clusters), and results heavily depend on such initial assumptions.

We divide the algorithms as metric and non-metric depending on their specific operating principles. Metric algorithms employ measures of similarities and distances to the predictors whereas non-metric algorithms do not. Among the metric-based algorithms, we consider Support Vector Machines \citep[SVM,][]{hearst1998support} which uses a kernel to compute the inner product of all pairs of data in the feature space and implicitly projects the data in a higher dimensional space where such data are linearly separable; and $K$-Nearest Neighbors \citep[KNN,][]{duda2012pattern}, which uses a suitable similarity function/distance to evaluate the closeness of a new sample to samples stored in memory. Among the non-metric algorithms, we consider Decision Trees \citep[DT,][]{safavian1991survey}, which construct a tree structure and explore nodes and leaves for both classification and regression; and Random Forest \citep{breiman2001random}, which is an ensemble of multiple DT, where each tree is constructed by sampling a random set of attributes from the data. Each tree performs regression via a mean prediction and classification via majority voting. Ensemble methods \citep[e.g., Bootstrap Aggregation or Bagging,][]{breiman1996bagging}, where an ensemble of weak learners are combined to produce a stronger learner, are considered for both regression and classification tasks.

\subsubsection{Classification task: Support Vector Machine}
\label{SVM}

In SPH, the continuous fluid is represented as a Lagrangian set of particles that move with the flow; this allows easy visualization and supports analytical deductions \citep[and references therein]{2015aste.book..661A}. We digest the dataset for the classifier by defining the qualitative outcome of each giant impact simulation according to four distinct classes of responses: merging, disruption, ``graze-and-merge'', ``hit-and-run'' \citep[e.g.][]{2006Natur.439..155A,2015aste.book..661A,Stewart2009}, Figure \ref{movie_nigth}. In our classification, we distinguish between merging and ``graze-and-merge'' scenarios. The latter is a transient evolution that would eventually lead the projectile to merge with the target, but escaping bodies up to a third of the size of the progenitors can be spin off during the collision \citep{2013Icar..223..544A}. The outcome of the simulations (response, or class) is associated to four impact parameters (predictors): mass of the target, projectile-to-target mass ratio, impact angle, and impact velocity. The dataset has entries:

\begin{equation}
    \{(M_\mathrm{T},~\gamma,~\theta_\mathrm{coll},~v_\mathrm{coll}/v_\mathrm{esc}) ; \mathrm{class}\}
\end{equation}
\noindent
where $M_\mathrm{T}$ is the mass of the target, $\gamma=M_\mathrm{P}/M_\mathrm{T}$ is the projectile-to-target mass ratio ($M_\mathrm{P}$ being the mass of the projectile), $\theta_\mathrm{coll}$ is the impact angle and $v_\mathrm{coll}$ is the collision velocity normalized to the mutual escape velocity $v_{esc}$ (Equation \ref{esc}). The matching between predictors and response is done during one of our ``movie days'': four co-authors watched short movie clips of the simulations and, based on this visualization, agreed on the outcome of the simulations (class). When dealing with a complicated problem, a group of experts with varied experience in the same area have a higher probability of reaching a satisfactory solution than a single expert \citep{baruque2010committee}. Labeling error, however, can still occur for several reasons, including subjectivity, data-entry error, or inadequacy of the information used to label each entry. Domains in which experts disagree are natural places for subjective labeling errors \citep[and references therein]{brodley1999identifying}. To mitigate mislabeling and its negative effect on the performance of the classifier, the labeling of the dataset is performed by the domain experts with a majority vote. Taking a majority over many hypotheses, all of which proposed by different experts, has the effect to reduce the random variability of the labels \citep{baruque2010committee}. More advanced approaches to labeling \citep[e.g., crowd-sourcing or weighted-voting,][]{Rodrigues2013} or to labeling error mitigation \citep[e.g., ensemble learning,][]{zhang2012ensemble} are also possible, but they are beyond the scope of this pilot study.

\begin{table*}
    \centering
    \begin{tabular}{|c|c|c|c|c|}
    \hline
        Target mass [$\mathrm{M_\oplus}$] & Mass ratio (projectile/target) & Impact angle & Impact velocity [$v_{esc}$] & Collision class\\ \hline
       $1$ & 0.70 & 89.5 & 1.30 & ``hit-and-run'' (flag: 1)\\
       $1$ & 0.70 & 89.5 & 1.05 &  ``graze-and-merge'' (flag: 2) \\ 
       $1$ & 0.70 & 22.5 & 1.00 & merging (flag: 3) \\
       $1$ & 0.70 & 22.5 & 4.00 & disruption (flag: 4) \\
       $10^{-1}$ & 0.70 &  30.0 & 1.50 & ``hit-and-run'' (flag: 1)\\ 
       $10^{-1}$ & 0.70 &  30.0 & 1.40 &  ``graze-and-merge'' (flag: 2) \\ 
       $10^{-1}$ & 0.70 & 22.5 & 1.00 & merging (flag: 3)\\
       $10^{-1}$ & 0.70 & 22.5 & 4.00 & disruption (flag: 4) \\
        \hline
    \end{tabular}
    \tablecomments{ Table \ref{tab:class} is published in its entirety in the machine-readable format. A portion is shown here for guidance regarding its form and content.}
    \caption{Excerpt of the labelled data for the classification task. The elements in columns first to fourth are the predictors (pre-impact conditions): $M_T \in [10^{-2},~1]~ \mathrm{M_\oplus}$; $\gamma=M_\mathrm{P}/M_\mathrm{T} \in [0.2,~0.7]$; $\theta_{coll} \in [0,~90]$; $v_{coll}/v_{esc} \in [1,~4]$. The elements in the fifth column are the responses (type of collision outcome). Among the responses, ``hit-and-run'' cases are coded as \#1; ``graze-and-merge'' cases are coded as \#2; merging cases are coded as \#3; disruptive cases are coded as \#4.}
    \label{tab:class}
\end{table*}

An excerpt of the labelled data is reported in Table \ref{tab:class}. The dataset for the classification task is published in its entirety in the machine-readable format. Among the available schemes, we selected a multi-class Support Vector Machine \citep[SVM,][]{hearst1998support} as the algorithm achieving the highest validation for the classification task (see Section~\ref{class}). SVMs were introduced by Boser, Guyon and Vapnik \citep{boser1992training}, and became very popular because of their large success at the handwritten digit recognition task. SVMs are machine learning algorithms that can discriminate between different classes given input data. They are considered primary examples of the so-called ``kernel methods''.

Consider a set of given training vectors $\mathbf{x}_i \in \mathbb{R}^n, i =1,....,l$ that belong to two classes, as well as a class indicator vector $\mathbf{y} \in \mathbb{R}^l$ such that $y_i \in [-1,1]$. The basic SVM algorithm solves the following primal optimization problem:

\begin{equation}
    min_{\mathbf{w},b,\eta}~~ \frac{1}{2}\mathbf{w}^T\mathbf{w}+C\sum_{i=1}^l\eta_i
\end{equation}
\noindent
subject to the following constraints:

\begin{equation}
    y_i(\mathbf{w}^T\phi(\mathbf{x}_i)+b)\geq 1 - \eta_i,~~
    \eta_i \geq 0
\end{equation}

Here, $\phi(\mathbf{x}_i)$ maps the training vectors $\mathbf{x}_i$ into a higher-dimensional space. $C \geq 0$ is the Tikhonov regularization parameter. Generally, the vector variable $\mathbf{w}$ lives in a high dimensional space. Thus, one equivalently solves the following dual problem:

\begin{equation}
    min_{\alpha}~\frac{1}{2}\mathbb{\alpha}^T Q\mathbb{\alpha} - \mathbf{e}^T\mathbb{\alpha}
\end{equation}
\noindent
subject to:

\begin{equation}
    \mathbf{y}^T\mathbb{\alpha}= 0,~~~~0\leq \alpha_i \leq 0,~~i=1,...,l
\end{equation}

Here, $\mathbf{e} = [1,....,1]^T$ is a vector comprising all ones, $Q$ is an $l\times l$ positive semi-definite matrix where $Q_{i,j} = y_iy_j K(\mathbf{x}_i,\mathbf{x}_i)$. The kernel function $K(\cdot,\cdot)$ is defined as $K(\mathbf{x}_i,\mathbf{x}_i) = \phi(\mathbf{x}_i)^T\phi(\mathbf{x}_i)$. After the optimization problem is solved via the primal-dual relationship, the optimal vector $\mathbf{w}$ satisfies the following relationship:

\begin{equation}
    \mathbf{w} = \sum_{i=1}^l y_i\alpha_i\phi(\mathbf{x}_i)
\end{equation}

Importantly, the decision (discriminative) function for the binary classification problem is mathematically described as:

\begin{equation}
    sgn(\mathbf{w}^T\phi(\mathbf{x})+b) = sgn(\sum_{i=1}^l y_i\alpha_i K(\mathbf{x}_i,\mathbf{x})+b)
\end{equation}

This formulation holds when the problem has nonlinear decision surfaces, as the input vector $\mathbf{x}$ is substituted by a properly selected mapping function $\phi$ that projects the training data into a suitable feature space \citep{2009Shashua}. The choice of the function $\phi$ is done using $k-$fold cross-validation, which subdivides the training set in $k$ subsets and train the classifier (i.e., solve the primal-dual optimization problem) using only $(k-1)$ subsets. The validation accuracy (i.e., percentage of correct classification) is computed -- after training -- on the $k-$th subset. The procedure is repeated several times; the average validation accuracy is used to compare different schemes with different hyperparameters (i.e., the value of $k$ and the function $\phi$). The model with the highest validation accuracy is adopted.

Once the SVM is trained and validated, its performance is assessed by means of a confusion matrix computed on a testing set. The confusion matrix shows the degree to which the classifier is confused when it makes predictions; each row represents the instances in a predicted class while each column represents the instances in an actual class \citep{Ting2010}. As an example, for the binary sub-problem of classification between ``graze-and-merge'' (GnM) and ``hit-and-run'' (HnR), the confusion matrix has the form:

\begin{table}[H]
    \centering
    \begin{tabular}{|c|c|c|c|}
    \hline
        & Actual: HnR & Actual: GnM\\ \hline
        Predicted: HnR & $a$ & $b$ \\
        Predicted: GnM & $c$ & $d$ \\
        \hline
    \end{tabular}
\end{table}

The diagonal elements are the instances of correct classifications, while the off-diagonal values account for misclassifications. In this example, the total number of actual ``hit-and-run'' events is `$a+c$'; after training, the SVM classifies correctly `$a$' events and misclassify `$c$' events as ``graze-and-merge''. The accuracy of the classifier is computed as the percentages of true positives (correct predictions) over total number of sample: 

\begin{equation}
\label{acc}
    AC~[\%]= \frac{a+d}{a+b+c+d}\times 100
\end{equation}

Classification problems with a number of classes greater than 2 are decomposed into multiple binary classification problems, according to different transformation techniques \citep[e.g., one-vs-one and one-vs-rest strategies,][]{Bishop2006}. The choice of a specific technique is also part of hyperparameter optimization.

\subsubsection{Regression task: Neural Networks}
\label{NN}

\begin{figure}
\centering
  \includegraphics[width=\linewidth]{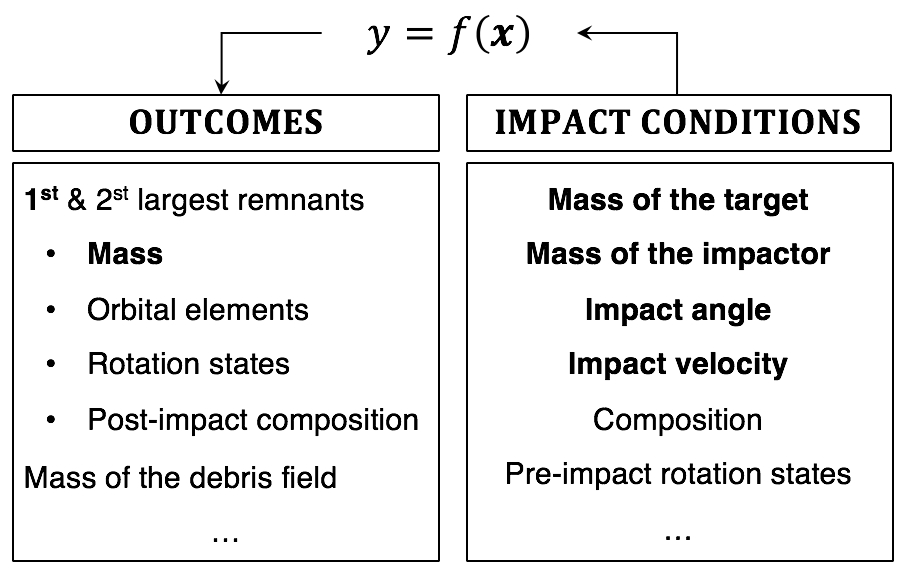}
  \caption{A surrogate model (e.g, neural network) is able to generalize the functional relationship $y=f(\boldsymbol{x})$ between  real-variable input $\boldsymbol{x}$ (impact conditions, right column) and outputs $y$ (collision outcomes, left column). Training occurs on $N$ data of the type: \{$\boldsymbol{x}$; $y$\}$_i$ = \{predictor; label\}$_i$, $i = 1,...,N$. Examples of impact conditions (predictors) and outcomes (labels) are shown in the right and left columns, respectively. In this pilot study, we train a neural network to associate four impact conditions (mass of the target, projectile-to-target mass ratio, impact angle, impact velocity) to accretion efficiency (or mass of the largest remnant, Equation~\ref{acceff}).}
  \label{generalization}
\end{figure}

Whereas classifiers are able to handle discrete, qualitative responses, a regressor is a surrogate model able to mimic the ``parent'' SPH input-output function to predict continuous (real-variable) outputs given the input parameters (predictors), Figure \ref{generalization}. This scheme provides a synthesis of the collision outcome in terms of a set of output properties of interests (e.g., mass of the largest remnants, their post-collision orbital elements, etc.), by learning from large planetary formation datasets of collision. Running the surrogate model drastically reduces the computational time with respect to full SPH simulations (from hours to seconds). We design a surrogate model for the prediction of accretion efficiency (i.e., the mass of the largest remnant of the collision) at several times the collision timescale, Equation~(\ref{tg}). After this time, pressure and temperature gradient forces are no longer acting and the resulting scenario (largest remnants and their orbital properties) can be treated using $N$-body integrator rather than hydrocodes. We use the definition of accretion efficiency by \citet{2010ChEG...70..199A}:

\begin{equation}
    \xi = \frac{(M_\mathrm{LR} - M_\mathrm{T})}{M_\mathrm{P}}
    \label{acceff}
\end{equation}
\noindent
where $M_\mathrm{LR}$ is the mass of the largest remnant, $M_\mathrm{T}$ is the mass of the target body and $M_\mathrm{P}$ is the mass of the projectile. For each simulation in our dataset, the largest remnants is identified as discussed in Section \ref{dataset}. A summary of the data is reported in Table \ref{tab:reg}. The dataset for the regression task is published in its entirety in the machine-readable format.

\begin{table*}
    \centering
    \begin{tabular}{|c|c|c|c|c|}
    \hline
        Target mass [$\mathrm{M_\oplus}$] & Mass ratio (projectile/target) & Impact angle & Impact velocity [$v_{esc}$] & Accretion efficiency (Eq. \ref{acceff}) \\ \hline
        $1$ & 0.70 & 52.5 & 1.15  & 0.02 \\
        $1$ & 0.70 & 22.5 & 3.00 & -0.58 \\
        $1$ & 0.70 & 45.0 & 1.30 &  0.02 \\
        $10^{-1}$ & 0.70 & 15.0 & 1.40  & 0.90 \\
        $10^{-1}$ & 0.20 & 15.0 & 3.50 & -1.52 \\
        $10^{-1}$ & 0.35 & 15.0 & 3.50 & -1.25 \\
        $10^{-2}$ & 0.70 & 60.0 & 1.70 & 0.00 \\
        \hline
    \end{tabular}
    \tablecomments{ Table \ref{tab:reg} is published in its entirety in the machine-readable format. A portion is shown here for guidance regarding its form and content.}
    \caption{Excerpt of the data for the regression task. The elements in columns first to fourth are the predictors (pre-impact conditions): $M_T \in [10^{-2},~1]~ \mathrm{M_\oplus}$; $\gamma=M_\mathrm{P}/M_\mathrm{T} \in [0.2,~0.7]$; $\theta_{coll} \in [0,~90]$; $v_{coll}/v_{esc} \in [1,~4]$. The elements in the fifth column are the responses (accretion efficiency $\xi$) as post-processed by \citet{2019ApJGabriel}.}
    \label{tab:reg}
\end{table*}

This effort of the work is entirely independent from the classification of Section~\ref{class}. For this task, a Neural Network (NN) is trained, validated and tested to replace the more computationally expensive ``parent'' numerical models, e.g., the full SPH simulation, in the prediction of accretion efficiency. NNs are able to learn (i.e., improve the performance of a specific tasks) from data, by modeling the functional relationship between inputs and outputs, which is exemplified by labeled data. NNs consist of many mathematical units called neurons, which communicate in a parallel fashion through weights that represent the strength of the corresponding synapses. Neurons are the basic processing units for the network and are characterized by an activation function $h(\cdot)$. Additive nodes with activation functions have the following structure:

\begin{equation}
     G(\boldsymbol{a}_i,b_i,\boldsymbol{x}) = h(\boldsymbol{a}_i^T\boldsymbol{x}+b_i)
\end{equation}
\noindent
where $\boldsymbol{a}_i \in \mathbb{R}^m$ and $b_i \in \mathbb{R}$. A common activation function for shallow neural networks is the tanh-sigmoid function

\begin{equation}
    h(s) = \frac{2}{1+ \exp(-2s)}-1
\end{equation}

For deeper architectures, such as Convolutional Neural Network \citep[CNN,][]{krizhevsky2012imagenet}, other activation functions (e.g., ReLu, Rectified Linear Unit) are more commonly used. Neurons are organized by layers; for this application, we adopt a shallow network comprising one input layer, one hidden layer and an output layer. The hidden layer is assumed to have a specified number of neurons $S$. The overall process begins with a summation of each input with the correspondent weights (synapses) and then further processing by an activation function. In regression problems, the overall NN output function is typically represented as follows:
\begin{equation}
    f_S(\boldsymbol{x}) = \sum_{i=1}^S\boldsymbol{\beta}_i h_i(\boldsymbol{x}) = \sum_{i=1}^S\boldsymbol{\beta}_i G(\boldsymbol{a}_i,b_i,\boldsymbol{x})
\end{equation}
\noindent
where $\boldsymbol{x}\in\ \mathbb{R}^d$ and $\boldsymbol{\beta}_i \in \mathbb{R}^m$. The weights $\boldsymbol{a}_i$ an biases $b_i$ are determined during the training process which implies minimization of a loss function. For regression problems, the typical loss function is the Mean Square Error (MSE), i.e.:
\begin{equation}
\label{MSE}
    MSE(\boldsymbol{a}_i,b_i) = \frac{1}{N}\sum_{i=1}^N(f_S(\boldsymbol{x}_i)-y_i)^2
\end{equation}
where $\{\boldsymbol{x}, y\}_i$ is the associated training set of size $N$. 

In this paper, the regressor is trained, validated and tested on data of the type:

\begin{equation}
\{(M_\mathrm{T},~\gamma,~\theta_\mathrm{coll},~v_\mathrm{coll}/v_\mathrm{esc});~\xi\}
\end{equation}
while the composition is kept as parameter. The dataset is subdivided in training, validation and testing subsets, typically in proportion  70\%-15\%-15\%. Network training is performed using the training set and involves the fitting of the network parameters (weight $\boldsymbol{a}_i$ and biases $b_i$) via minimization of the loss function (Equation \ref{MSE}). Common approach to training involves backpropagation or Stochastic Gradient Descent  \citep{schmidhuber2015deep}. A training step (or epoch) consists in a round of predictions for the predictors $\boldsymbol{x}_i$ in the training set, followed by backpropagation of the residuals between targets and prediction and update of weights $\boldsymbol{a}_i$ that reduce the MSE. At each epoch, the performance of the network (and progress toward a successful training) is evaluated in terms of Mean Square Error, MSE, Equation~\ref{MSE}, which is expected to decrease as the number of epochs increases, thus indicating progressive improvement in the performance (i.e., learning).  

Both the validation and testing sets are employed to protect against overfitting of the training set \citep{bishop1995neural}. The training process is not a simple interpolation of the training set, but rather it involves the search for families of parametric functions  (i.e.,  the  neural  network) that globally fit the data (generalization). The validation procedure consists in the search of those network hyperparameters (e.g., the learning rate or numbers of hidden neurons, which are not learned during training) that minimize the MSE on the validation set. In addition to help protecting against overfitting, the testing set is used for an independent assessment of the generalization capabilities of the network, i.e., the behavior of the MSE on an unseen ensemble of data. Properly trained networks ensure that the data in the validation and the testing sets follow the same probability distribution of the data in the training set. At every training epoch, the MSE for validation and testing is computed; the training is completed when the MSE on the validation set does not further decrease for 6 consecutive training epochs.

In addition to the MSE, the overall process is evaluated also in terms of regression value R, which measures the degree of correlation between outputs and targets; this quantity is the analogous of the SVM classification accuracy for real-variable data. The regression value R is a non-dimensional quantity and allows comparing the performance of different approaches to the problem (e.g., data-driven approach versus data interpolation) with respect to the data, at testing. An optimal result shows low MSE values (i.e., close to zero) and a high degree of correlation between predictions and targets (i.e., a R value close to 100\%) on the testing set.

\section{Results}
\label{sec:result}

In the following two sections, the trained response functions (classifier of collision types and regressor of accretion efficiency) are presented. We also discuss their predicting performance with respect to the labels of the entries in the datasets. 

\begin{figure*}
\centering
  \includegraphics[width=\linewidth]{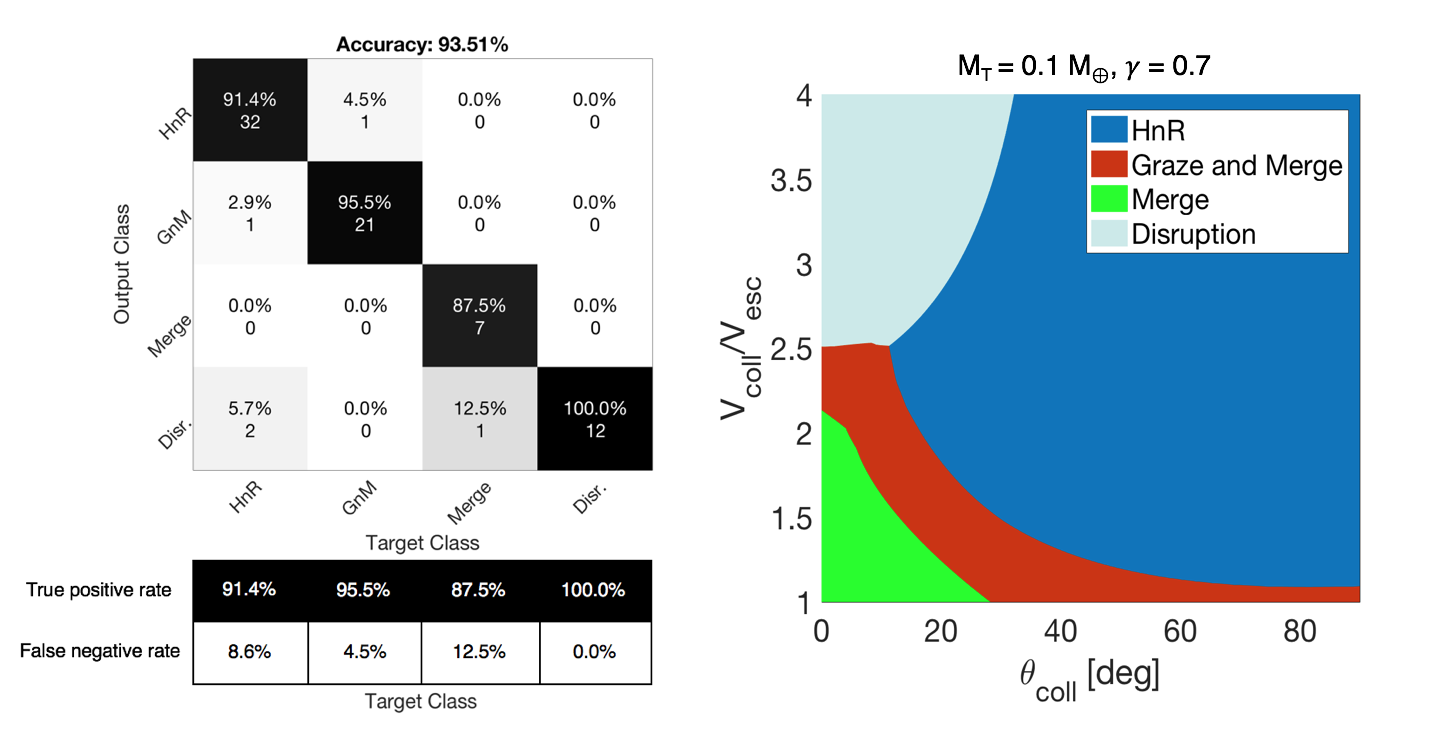}
  \caption{Left panel: Confusion matrix of the 4-D classifier, quantifying the degree of accuracy of the classification on the testing set. The elements on the diagonal of the confusion matrix represents those instances that have been correctly classified by the SVM (true positives). Conversely, each extra-diagonal element represents the number of mis-classifications with respect the SPH data (i.e., the labels). The number of misclassifications is added along each column to compute the false negative rates. Overall, we achieve a true positive rate of 91.4\% on the ``hit-and-run'' (HnR) class, 95.5\% for ``graze-and-merge'' (GnM) class, 87.5\% for the merge class and 100.0\% for the disruption class. The confusion matrix is close to be fully diagonal; the accuracy -- computed as the mean value of the true positives over the whole population, Equation~(\ref{acc}) -- is above 93\%. Right panel: decision boundaries for the collision type, as predicted by the classifier for a mass of the target $M_\mathrm{T}=\SI{0.1}{\mearth}$ and a mass ratio between the projectile and the target $\gamma=0.7$. The impact velocity spans in a range between 1 to 4 times the mutual escape velocity (Equation \ref{esc}) while the impact angle ranges from head-on to grazing configurations.}
  \label{class_ex}
\end{figure*}

\begin{figure*}
\centering
  \includegraphics[width=\linewidth]{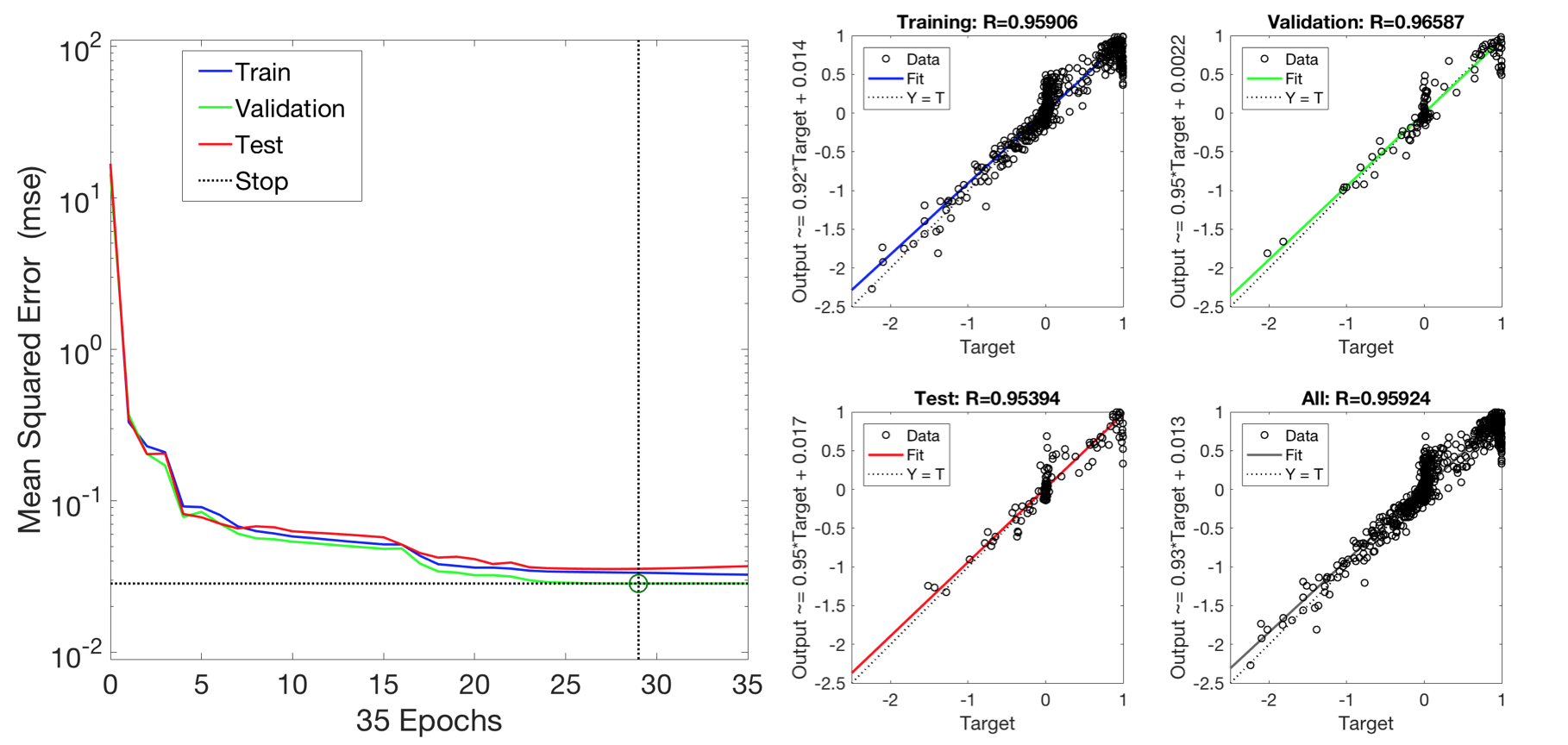}
      \caption{Left panel: evolution of the Mean Square Error (MSE) for training, testing and validation, for increasing epochs of training. When validation is concluded, the average plateau value of the testing MSE is 0.04. This quantifies the global uncertainty of the surrogate model in mimicking the ``parent'' numerical model, i.e., the SPH simulations. Right panel: correlation between predictions and target, and overall fitting with respect to an expected 1:1 line. The regression index R is about 96\% (average), close to the optimal value of 100\%.}
  \label{NNperf}
\end{figure*} 

\subsection{Classifier of collision outcomes}
\label{class}

The classifier of collision outcomes maps the four impact properties (mass of the target, projectile-to-target mass ratio, impact angle, impact velocity) into one of the following types of collisions: merging, disruption, ``hit-and-run'', ``graze-and-merge''. The classifier is trained, cross-validated and tested as discussed in Section~\ref{SVM}. The ensemble of 769 labelled SPH simulations in Table \ref{tab:class} is split in a training dataset (90\%) and a testing dataset (10\%) via random sampling without replacement. The training set is used for training the network with 10-fold cross-validation, which allows performing hyperparameter optimization for what concerns the best kernel feature parametrization. We find that a quadratic kernel ($K=\phi^T \phi = (k^T x + m)^2$) achieves the best cross-validation accuracy (91.0\%).

The performance of the classifier, in terms of its confusion matrix, is shown in Figure \ref{class_ex}, left panel. The performance is evaluated on the testing set, corresponding to 77 entries, which was not used for training and cross-validation. Testing the algorithm on this separate dataset provides an independent, additional assessment of the performance of the classifier on unseen data. We achieve an overall accuracy above 93\% at testing. Certain regimes, however, are characterized by more misclassifications (e.g., disruption versus merging) than others (e.g., ``hit-and-run''). Those classes characterized by high false negative rates prevent the classifier from achieving 100\% accuracy at testing (i.e., a fully diagonal confusion matrix); this is found to be indicative of ``confusion'' along the decision boundaries between regimes, as we address in more details in Section~\ref{guide} (Figure \ref{score}, left panel).

The classifier is intrinsically a 4-D scheme, with many dimensions as the number of predictors (impact properties). The algorithm describes the outcome in parameter space by means of decision hyper-surfaces, which mark the transition between different regimes. To better appreciate these features, the parameter space can be sectioned in 2-D slices; one example of these map is in Figure \ref{class_ex}, right panel, for a mass of the target $M_\mathrm{T}=\SI{0.1}{\mearth}$ and similar-mass projectile ($\gamma=M_\mathrm{P}/M_\mathrm{T}=0.7$). The collision type is mapped into a space of collision velocity (in units of mutual escape velocity) and impact angle. We recognize 4 distinct collision regimes, whose decision boundaries are the traces of the decision hyper-surfaces suggested by the classifier. Each regime is a ``phase'' in which the collision outcome is qualitative similar, that is, a scaling law is expected to apply.

\subsection{Regressor of accretion efficiency}
\label{surr_mod}

The neural network has 4 input neurons (as many as the impact properties), one hidden layer, and one output layer which predicts accretion efficiency. The dataset of Table \ref{tab:reg} is composed by 810 simulations; their predictors (i.e. the impact properties) are internally scaled in a min-max procedure. The training is performed using the Levemberg - Marquardt algorithm \citep{demuth2014neural} on 70\% of the overall dataset. The rest of the data is split between a validation set (15\%) and a testing set (15\%). The dataset is split via random sampling without replacement to assure that the data in the three sets follow the same probability distribution. Figure \ref{NNperf}, left panel, shows learning dynamics in terms of the evolution of the MSE for training, validation and testing, at different epochs of training procedure. For the hidden layer, the choice of 10 hidden neurons gives the lowest MSE at validation. The testing MSE converges to an error level of about 0.04. This value is an estimate of the global accretion efficiency error, as it quantifies the (squared) residual between the values predicted by the regressor and the values of accretion efficiency of the SPH data in the testing set. The training error is also 0.04, while the validation error is about 0.03. Figure \ref{NNperf}, right panel, shows the correlation index at the end of the training procedure, whose value is above 95\% on testing. 

As for the classifier of collision outcome, the regressor maps accretion efficiency in a 4-D parameter space. For a mass of the target $M_\mathrm{T} = \SI{0.1}{\mearth}$ and similar-mass projectile ($\gamma=M_\mathrm{P}/M_\mathrm{T}=0.7$), Figure \ref{comp_SL}, left panel, shows a 2-D map (slice of the parameter space) of accretion efficiency in a plane of impact velocity (in units of mutual escape velocity) and impact angle. The grid has a step of $0.01^\circ$ along the impact angle axis ($\theta_{coll}$) and $0.01$ along the velocity axis ($v_{coll}/v_{esc}$). Accretion efficiency is color-coded such that the outcome varies from perfect merging (dark blue) to partial accretion (light blue) to partial erosion to disruption (redder colors and black for $\xi\leq-1$). The corner of the face that has the smallest indices determines the constant color of each mesh face. Catastrophic disruption is achieved when the mass of the largest remnant is less or equal to the half of the total mass of the system $(M_\mathrm{T} + M_\mathrm{P} )$. Given that $M_\mathrm{P} =\gamma M_\mathrm{T} $, catastrophic disruption is characterized by an accretion efficiency equal or less than $\xi_D = 0.5 - 0.5~\gamma$, with disruption threshold ($\xi_D = -0.21$ for $\gamma = 0.7$).

\section{Discussion}
\label{discussion}

\begin{figure*}
\centering
  \includegraphics[width=\linewidth]{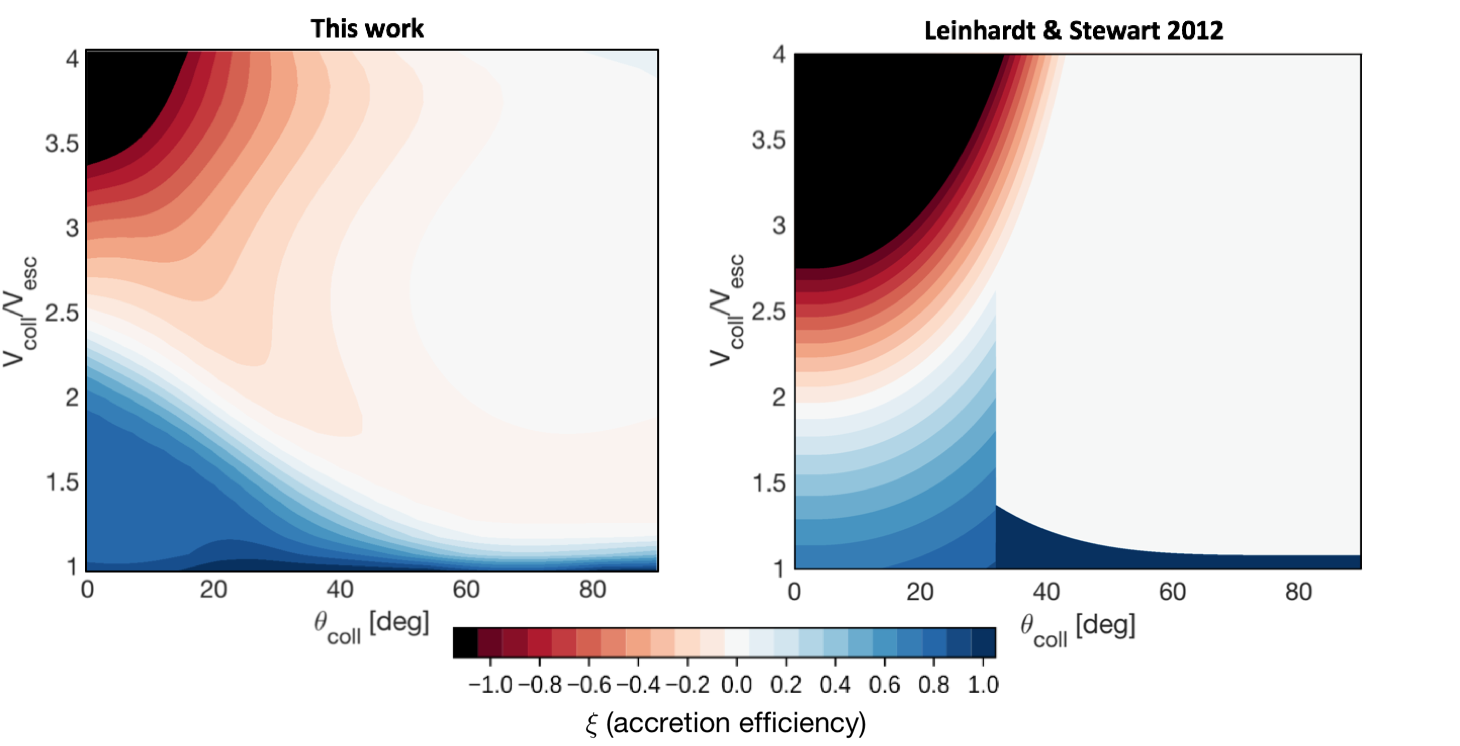}
  \caption{Left panel: map of accretion efficiency -- Equation~(\ref{acceff}) -- as predicted by the neural network (Section~\ref{surr_mod}). Right panel: map of collision outcome and accretion efficiency generated using the scaling laws proposed by \citet{2012ApJ...751...32S}, for the same combination of mass of the target and mass of the projectile, using the values $c^\star=1.9$ and $\bar{\mu}=0.36$, which  were fit to hydrodynamic planets}. Impact velocity (y axis) ranges between 1 to 4 $v_{esc}$, impact angle (x axis) ranges from head-on to grazing, $M_\mathrm{T} = \SI{0.1}{\mearth}$, and $\gamma = M_\mathrm{P}/M_\mathrm{T} = 0.7$. The grid was sampled in steps of $0.01^\circ$ and $0.01v_{\rm{esc}}$; color for each mesh face is dictated by the vertex with the smallest index. Accretion efficiency shows a rich range of outcomes, which includes transitions from accretion (cooler colors) to disruption (warmer/black colors), to hit-and-run (almost net-zero accretion; white colors).
  \label{comp_SL}
\end{figure*}

High-resolution SPH simulations have been used to train, validate and test a classifier of collision type (Section~\ref{class}) and a regressor of accretion efficiency (Section~\ref{surr_mod}). Together with the prediction of the type of collision (e.g., merging versus disruption), real-variable collision outcomes (e.g., mass of the larges remnants, their post-collision orbits) are needed to realistically simulate collisions in an N-body dynamical evolution. The regression of such quantities can be done by means of a neural network able to map pre-impact conditions into outcomes (Figure \ref{generalization}). In this work we present a first machine-learned regressor that predicts the accretion efficiency at many times the collision timescale -- Equation~(\ref{tg}). 

The two surrogate collision models (classifier and regressor) describe a 4-D parameter space in terms of mass of the target, projectile-to-target mass ratio, impact velocity and impact angle.  Interpretation by the human operator, however, is preferably done on a 2-D section (``slice'') of the 4-D parameter space. As an example, SVM decision boundaries and accretion efficiency are predicted for $M_\mathrm{T}=\SI{0.1}{\mearth}$ and $\gamma = M_\mathrm{P}/M_\mathrm{T} = 0.7$ in Figure~\ref{class_ex}, right panel, and Figure~\ref{comp_SL}, left panel, respectively. The map of the accretion efficiency unveils a richer background scenario in regions were the collision outcome seemed homogeneous according to the classifier. ``Graze-and-merge'' and merging are somewhat inefficient in delivering mass to the target, as a portion of the projectile as high as 50\% can escape accretion. Because of its typical grazing nature, the ``hit-and-run'' regime is characterized by accretion efficiency close to 0, within the error associated with the training. At the most probable impact angle \citep[i.e., 45$^\circ$,][]{1962BookShoemaker}, however, lower-energy ``hit-and-run'' cases are indistinguishable from partially-accreting ``graze-and-merge'' events, while partial erosion starts to dominate above $v_\mathrm{coll}/v_\mathrm{esc}\sim 2$. Overall, the target is likely to be slightly eroded in the ``hit-and-run'' regime, but the second largest remnant (i.e., the surviving projectile) underwent the highest collision and tidal stresses, as the energy of the impact is partitioned equally in the two bodies.

\subsection{Comparison with scaling laws}

Predicting the outcome of a giant impact without performing a full hydrodynamics simulation has already been the subject of multiple studies leading to the formulation of scaling laws \citep[e.g.,][]{1990Davis,Benz1999,2012ApJ...745...79L}. A scaling law is an analytic relationship between impact properties (e.g., mass ratio, impact angle, and impact velocity) and its outcome for any collision in a physical regime (e.g, between gravity-dominated bodies), assuming invariance with respect to one property, usually the mass of the target. Hydrodynamical simulations are used to fit the parameters of the relationship, and, ideally, account for the transition between the different regimes. 

Here, we compare our results with one such law by \citet{2012ApJ...745...79L}. They proposed scaling the collisions according to the ratio between the specific impact energy, and the catastrophic disruption threshold $Q^*_{RD}$ -- the specific energy required to disperse half the total colliding mass (for non-grazing collision). The reference specific energy is first computed for head-on collisions between equal-mass bodies and then corrected for the mass ratio and impact angle.

The left panel of Figure~\ref{comp_SL} shows the map of accretion efficiency (predicted using our regressor), again for $M_\mathrm{T}=\SI{0.1}{\mearth}$ and $\gamma = 0.7$. On the right panel of Figure~\ref{comp_SL} is the analogous map generated using the scaling laws for hydrodynamic bodies proposed by \citet{2012ApJ...745...79L}. The fit parameters in their model which are most relevant to our results are $c^\star=1.9\pm0.3$ and $\bar{\mu}=0.36\pm0.01$ and were thus used to generate the right panel of Figure~\ref{comp_SL}. Our data-driven approach and the empirical, physics-based energy scaling by \citet{2012ApJ...745...79L}, however, are different in two fundamental aspects: 1) the underlying dataset of simulations that were used for fitting procedures; and 2) the fitting methodology. Because of these differences, we keep the comparison between the two results shown in Figure~\ref{comp_SL} qualitative and aim to highlight similarities and differences between those.

\citet{2012ApJ...745...79L} segregate collisions into ``grazing'' and ``non-grazing'' according to the critical impact parameter $b_\mathrm{crit}=\sin{\theta_\mathrm{crit}}=R_\mathrm{T}/(R_\mathrm{T}+R_\mathrm{P})$ \citep{2010ChEG...70..199A} (see vertical line in the right panel of Figure 6). This relationship, however, was introduced by \citet{2010ChEG...70..199A} as a geometrical guideline and not intended for the purpose of accurately predicting ``hit-and-run'' events. The description of the parameter space by our surrogate models does not show a ``hard'' transition between grazing and non-grazing scenarios based on the critical impact parameter value. We unveil the occurrence of ``hit-and-run'' events at angles lower than the critical value, discussed further in \citet{2019ApJGabriel}.

In the grazing domain (on the right of the critical impact angle), \citet{2012ApJ...745...79L} assume that all collisions are ``hit-and-run'' in nature for sufficiently high impact velocities and accretion efficiency is assumed to be zero, i.e., the largest and second largest remnant masses are equal to the target and projectile mass respectively. In the ``hit-and-run'' regime, we confirm that the accretion efficiency is consistently close to zero (within the accuracy of the regressor) in the majority of the parameter space, but partial accretion or erosion scenarios are recorded close to transition with other regimes (Figure \ref{comp_SL}, left panel).

Our surrogate models show that perfect merging is rare; it may happen for low impact velocities and mid impact angles (about \SI{15}{\degree} to \SI{50}{\degree}, again, within the accuracy of the regressor). Grazing events need to eject some material to release angular momentum that would otherwise lead to unphysical spin \citep{2013Icar..223..544A}. Most of the regions categorized by the classifier as merging or ``graze-and-merge'' is actually partial accretion rather than perfect merging. The underlying events were categorized as such since the lost mass is in the form of debris.

At the boundary between the ``hit-and-run'' and ``graze-and-merge'' regimes (low impact velocity and high impact angle), the transition curve by our classifier of collision outcome (decision boundary in Figure \ref{class_ex}, right panel) is found to be similar to that by \citet{2012ApJ...745...79L}, who use the hit and run velocity criterion from  \citet{kokubo2010} to mark the transition. Across this region, however, we also observe a rapid decrease in accretion efficiency -- from merging to ``hit-and-run'' values -- as the impact velocity increases (Figure \ref{comp_SL}, left panel).
 
We point out the similarity between the transition curves from our classifier (Figure~\ref{class_ex}, right panel) and that of \citet{2012ApJ...745...79L} (Figure~\ref{comp_SL}, right panel) at the boundary between the ``hit-and-run'' and the partial erosion regimes. For non-grazing scenarios, \citet{2012ApJ...745...79L} determine the outcome by specific impact energy and $\gamma$ solely. Accretion efficiency ranges from partial accretion (cool colors in Figure~\ref{comp_SL} to catastrophic disruption (black color); catastrophic disruption for this combination of parameters is $\xi\leq -0.21$. Besides the differences in the assumed boundaries between regimes, our simulations are based on different underlying datasets. Our data-driven model is based on simulations from \citet{2011PhDReufer}, whereas the hydrocode simulations used in \citet{2012ApJ...745...79L} are from diverse source models \citep[e.g.,][]{2007SSRv..132..189B,2009Marcus,2010MarcusB}. \citet{2019ApJGabriel} demonstrate that the range of disruption thresholds exhibited by our dataset are close to the uncertainty of the disruption threshold of \citet{2012ApJ...745...79L}. Thus, we do not consider the difference in disruptive behavior observed in Figure~\ref{comp_SL} to be significant.

\subsection{A guide to parameter space exploration}
\label{guide}

\begin{figure*}
\begin{center}
  \includegraphics[width=\linewidth]{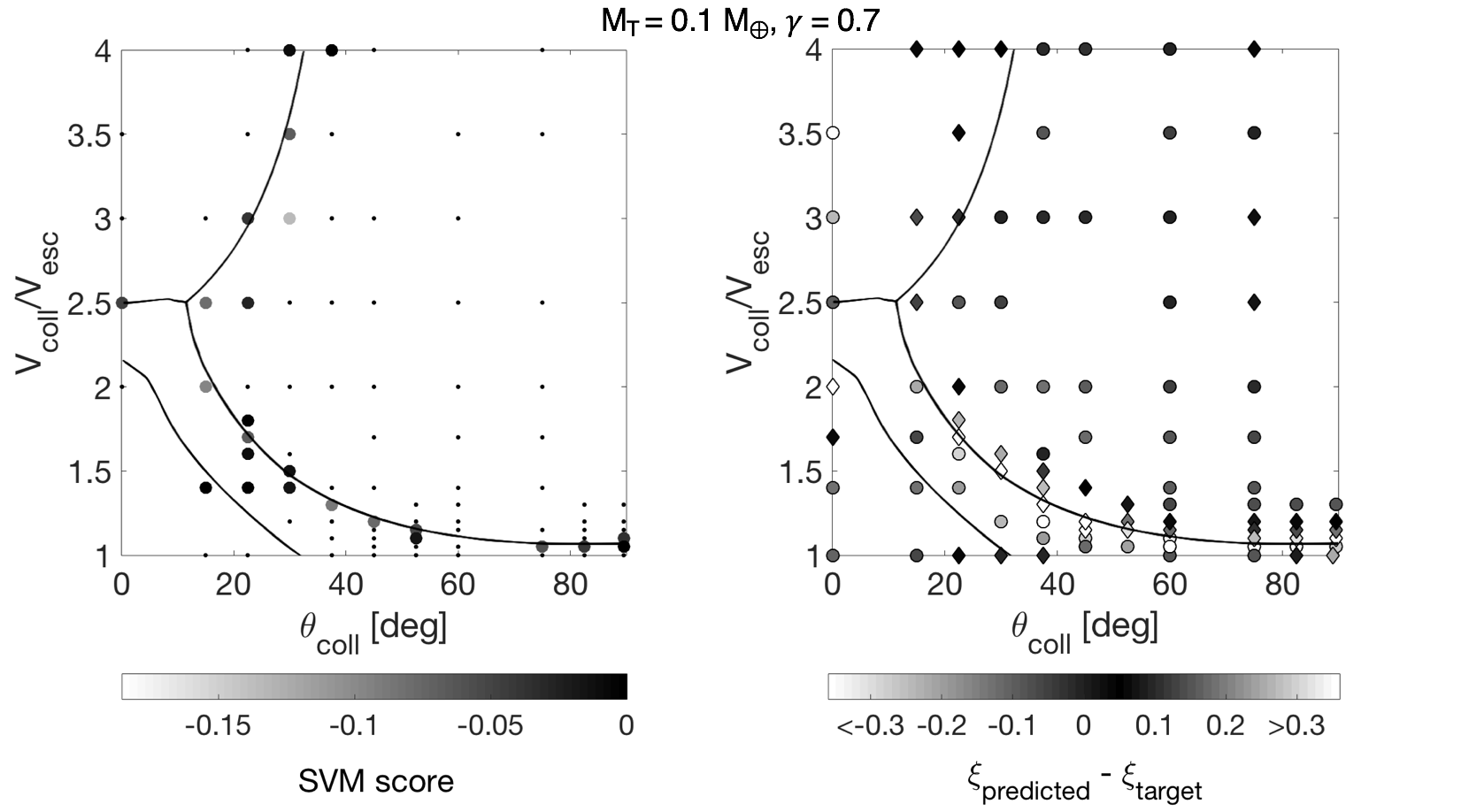}
  \caption{Left panel: local SVM score of the SPH simulations by the classifier of collision outcomes (small dots: correct predictions; color-coded datapoints: misclassifications). Right panel: residuals between the predictions by the regressor and the SPH data, for the same combination of mass of the target and mass of the projectile (diamonds: positive residuals; dots: negative residuals). Impact velocity ranges between 1 to 4 times the mutual escape velocity, impact angle ranges from head-on to grazing, $M_\mathrm{T}=\SI{0.1}{\mearth}$ and $\gamma=M_\mathrm{P}/M_\mathrm{T}=0.7$. High uncertainty is recorded along the decision boundaries (black curves), where misclassifications and inaccurate predictions tend to cluster. In these regions, additional SPH simulations are required to further reduce the ``confusion'' of the machine learning algorithms.}
  \label{score}
\end{center}
\end{figure*}

The designed classifier and regressor have high global accuracies (Figure \ref{class_ex}, left panel and Figure \ref{NNperf}), but misclassifications and inaccurate predictions can still occur locally in the parameter space. For a classifier, the local degree of confusion is quantified by the SVM classification score, which is the signed distance to the decision boundary. If the classifier is asked to predict the class for a labeled data, a positive, large score on the correct label means that the prediction is correct (the data are within the sub-space of the correct class), while a negative score indicates misclassification; the more negative the value, the higher the signed distance from the decision hyper-surface. The decision boundaries -- transition curve from a collision regime to another -- are in regions where the score tends to be negative, as the outcome is more sensitive to slight variations in the pre-impact conditions and mislabeling is likely to occur. This is evident in Figure \ref{score}, left panel, which shows the classification scores for the data with $M_\mathrm{T}=\SI{0.1}{\mearth}$ and similar-mass projectile ($\gamma=M_\mathrm{P}/M_\mathrm{T}=0.7$), in a plane of impact velocity and impact angle. Correct predictions are represented using small dots, while misclassified datapoints are color-coded according to their score (i.e., signed distance from the true classification boundary). The larger the absolute value of the score, the more severe the misclassification. The decision boundaries from the classifier are also reported (black curves). As expected, misclassification occurs more often along the boundaries. 

A similar trend is observed in Figure \ref{score}, right panel, where the predicted values for accretion efficiency are locally compared directly to the SPH data, again for $M_\mathrm{T}=\SI{0.1}{\mearth}$ and similar-mass projectile ($\gamma=M_\mathrm{P}/M_\mathrm{T}=0.7$), in a plane of impact velocity and impact angle. For the regressor, the local accuracy is quantified in terms of the residuals between predictions and targets (geometric distance). Inaccurate predictions are ``more distant'' with respect to their corresponding SPH data than accurate predictions. A positive (negative) residual indicates that the regressor is overestimating (underestimating) accretion efficiency with respect to the target value. In Figure~\ref{score}, predictions with positive residuals are represented using diamonds, while predictions with negative residuals are represented using dots. For the whole datasets (810 entries) 49\% of the predictions have positive residuals and the remaining 51\% cases have negative values. Therefore, the regressor is found to not systematically overestimate or underestimate accretion efficiency. We note that inaccurate predictions occur near the decision boundaries (black curves), which is to be expected. Local accuracies are thus expected to vary depending on location in the parameter space. The residual distribution is well approximated by a Gaussian fit centered at zero with 1-$\sigma$ value equal to 0.18. Large areas are characterized by residuals $<0.1$, and few cases (less than 1\%) have residuals up to 0.66 near transition regimes (absolute value, accretion efficiency units).

The distributions of SVM scores (local uncertainties for the classifier) and residuals (local uncertainties for the regressor) provide a guideline towards the completeness of the dataset, by indicating those regions of the parameter space that require additional simulations. The decision boundaries must not be intended as binary hard boundaries between different regimes of giant impact outcome, but rather as indicators that the outcome is gradually transitioning from a regime to another. Examples are the transition regions between ``graze-and-merge'' and merging (low impact angle and low impact velocity, left bottom corner of the panels in Figure \ref{score}), and ``hit-and-run'' and ``graze-and-merge'' (high impact angle and low impact velocity, right bottom corner of the panels in Figure \ref{score}). 

The extent of the transition regions is given by the size of the clusters of inaccurate classifications and predictions (``uncertainty band''). For the classifier, the uncertainty band quantifies the degree of ``confusion'' of the field experts during the labeling process. Such confusion arises because, near and along the decision boundaries, the outcomes of collision events seem alike or are unclear to the experts performing the labeling. These cases include the distinction between impactor disruption \citep[e.g.,][]{Leinhardt2012} and ``hit-and-run''. Furthermore, in proximity of certain decision boundaries, the outcome of a collision is highly sensitive to small changes in the impact parameters. For this reason, misclassifications correlate with inaccurate predictions by the regressor in the transition regions. Accretion efficiency is a real-number physical quantity and its transitions are smooth due to the occurrence of runner disruption at the boundary between erosive and ``hit-and-run'' collisions. The local gradient, however, can be large and more simulations may be needed for the regressor to resolve the region, i.e., to accurately learn the functional relationship between pre-impact conditions and accretion efficiency along the decision boundaries.

On the other hand, in regions where classification is exact and regression is accurate, one can avoid running a full SPH simulation to figure out the outcome, because the classifier is certain in the prediction of the type of collision, and the regressor is able to mimic the ``parent'' model at high fidelity.

\section{Conclusion and future work}
\label{concl}

We have applied machine learning to explore a large dataset of SPH simulations for giant impacts \citep{2011PhDReufer,2019ApJGabriel}. The relationship between beginning state (e.g., target mass, projectile mass, impact velocity and impact angle) and end state (impact outcome) has been mapped using two approaches. The result is the prototype of a full surrogate model of planet-forming giant impacts, which does not suffer from assumed physical models, and run in a fraction of a second, compared to days of simulation effort, thus enabling a fine -- and fast -- mapping of the parameter space to a known level of accuracy. 

First, we train, validate and test a Support Vector Machine \citep[SVM,][]{hearst1998support} to predict the type of the collision among 4 classes: merger, ``graze-and-merge'', ``hit-and-run'', and disruption. The classifier has global accuracy above 93\% at testing (Figure \ref{class_ex}, left panel), but local misclassifications are found to occur in proximity of the decision boundaries (Figure \ref{score}, left panel). Second, we train a neural network to predict the accretion efficiency, i.e., mass of the largest remnant of the collision. The network has a global error level of 0.04 (Mean Square Error between predictions and the dataset of SPH accretion efficiencies) and regression index above 95\% at testing (left and right panel in Figure~\ref{NNperf}, respectively); locally in the parameter space, residuals can reach 0.66 in accretion efficiency units (absolute value) but are generally lower, depending on the parameter region (Figure~\ref{score}, right panel). These functions -- classifier of collision outcome and regressor of accretion efficiency -- are called ``surrogate models'' because they provide a synthesis of the collision outcomes without the need to run a full hydrodynamical simulation. They are derived by generalizing the functional relationship between impact properties and outcomes, derived from the SPH simulations, to the whole parameter space within the ranges of the dataset (Table~\ref{tab:dataset} and Figure~\ref{histo_data}). The use of surrogate models avoids performing additional simulations over the entirety of the parameter space, which would be computationally inefficient given the large number of parameters and the requirement for high-resolution simulations to produce reliable outcomes. 

The present training has been done using a dataset that is sparse in many regions of importance. One feature of machine learning is that the surrogate models can be easily updated if the training landscape is expanded as new simulations become available. Future collision surrogate models will benefit from the publication of datasets available to researchers in the community. A proposed list of impact conditions and correspondent collision outcomes for use in realistic N-body dynamical studies of planetary formation can be found in Figure \ref{generalization}. Additional interesting outcomes include the thermodynamic history of the hydro-particles (pressure, temperature, and density) which provides insights into the composition and size distribution of the debris field.

For the present work we have trained on giant impacts in the gravity regime, where material strength plays a negligible role in the mass of post-impact remnants. In future work we shall extend the parameter space to ``small giant impacts'' involving bodies hundreds to thousands of kilometers diameter, colliding at around their mutual escape velocities, hundreds to thousands of meters per second. In this regime, friction plays a non-negligible role \citep[e.g.,][]{2015P&SS..107....3J,2017plan.book....7A}. New inroads have been made into SPH modeling of friction-governed planetary collisions \citep{2015P&SS..107....3J,2018Icar..301..247E,2018Sugiura} revealing its importance in thousand-kilometer-scale (embryo-embryo) collisions. Collisions in the friction regime have also been studied using soft-sphere discrete element (DEM) contacts in code PKDGRAV \citep{Schwartz2012} that has been applied to asteroid family formation \citep{Michel2001,Michel2004}, ejecta cloud evolution \citep{Schwartz2016}, and comet formation through catastrophic disruption \citep{Schwartz2018}. Angle of internal friction and material composition  \citep[e.g., icy versus chondritic,][]{Schwartz2018} are found to have a significant effect on the mass of the largest remnant \citep{Ballouz2014, Ballouz2015}. On asteroids, intergranular cohesion \citep{Scheeres2010} becomes a sizeable source of tensile strength, which may affect the impact outcome. Resolving this complex physics requires higher numerical resolution and much more computational overhead per timestep of evolution. At the larger extreme, there are few sets of data regarding giant impacts for planets larger than the Earth \citep[see][]{2009Marcus,2010MarcusA,2010MarcusB,2015ApJ...812..164L, 2018ApJKegerreis}; here a primary challenge is the reliable treatment of massive atmospheres. The same techniques of surrogate model development can be applied to these simulations, ultimately forming a general surrogate model for similar-sized planetary collisions at every scale, but, to date, no data table has been published at every scale.

The surrogate model is only as good as the post-processing of the physical simulations that gives us the derived outcomes for each run. The masses of the final bound remnants, and their velocities, rotations and compositions, must be reliably determined. In this study, the final masses have been computed using a friends-of-friends analysis and a calculation of binding energy; this is an approximation compared to running the simulation out many days longer in time to get the final bound objects, which then is increasingly effected by inaccuracies in the integrator. The application of Convolutional Neural Networks \citep{krizhevsky2012imagenet} could also improve significantly the reliability of clump detection, allowing for a more accurate identification and classification of second- and third-mass planets or planetesimals emerging from accretion-regime giant impacts. If it is possible to reliably identify bound clumps much earlier in a calculation, then emphasis could be placed on higher numerical resolution rather than longer runtime.

The combination of giant impact studies and machine learning is new research, and we anticipate many future studies \citep[e.g.,][]{2019valencia}. Machine classification is able to ``corral the herd'' of thousands of high-resolution simulations to identify the underlying structure of the parameter space. Machine regression is able to produce a quick and efficient algorithm for accretion efficiency that can be used in dynamical models, e.g., \textit{N}-body codes studying the growth of planets. These constitute the prototype of a surrogate model, that will reliably map inputs to outcomes and will effectively be equivalent to running an SPH simulation as an intermediate step during \textit{N}-body studies of planet formation. In fact, the surrogate models may become preferred since they run on a expediant functional call, yet are trained on high-resolution simulations instead of low-resolution simulations that would be run `on-the-fly'.

Because it represents simulation-derived data as a \textit{function}, a surrogate model can be inverted to formally understand the likelihood of specified scenarios of planet formation, such as Theia deriving from nearby the Earth, or Mercury forming in a couple of ``hit-and-run'' collisions \citep{2018ChauMercury}. Such inversion can be performed by means of Markov Chain Monte Carlo Bayesian inference \citep{stuart2010inverse} of observed post-collision scenarios, in which the surrogate models are used to sample the (unknown) posterior distribution of pre-impact conditions. Recent uses of this approach in planetary science include a new technique for constraining the thermal inertias of rock and regolith, and relative rock abundance, on asteroids from observed infrared fluxes \citep{cambioni2019constraining}. Rather than a boutique of scenarios that can solve for the origin of a given planet, there can be an inversion of outcomes. Lastly, there is an unknown future significance of machine learning in studies of planet formation, where \textit{unsupervised} classification of these datasets can reveal new and unforeseen trends and relationships in the data, leading to the development of better scientific models. Humans are excellent at looking for patterns in 2-D and 3-D datasets, but $N$-dimensional trends can often be performed better by a computer, leading to accurate data-driven models and scaling laws that help explain why collisions happen the way they do.

\acknowledgements

S.C., E.A., A.E. and S.R.S. acknowledge support from NASA Planetary Science Division (PSD) -- grant title ``Application of Machine Learning to Giant Impact Studies of Planet Formation'' -- and the University of Arizona. The authors thank the anonymous referees for the precious comments and suggestions that improved this manuscript.

\bibliographystyle{aasjournal}
\bibliography{ref,ref2,ref3,ML}

\end{document}